\documentclass[review,12pt]{elsarticle}

\usepackage{amssymb}
\usepackage{amsthm}
\usepackage{framed} 
\usepackage{amsmath,color}
\usepackage{mathrsfs}
\usepackage{graphicx}
\usepackage{epstopdf}
\usepackage{float}
\usepackage{caption}
\usepackage{subcaption}
\usepackage{bm} 
\usepackage{bbm}
\usepackage{mathrsfs}
\usepackage{cleveref}
\usepackage{soul}
\usepackage{accents}
\usepackage{color,soul} 
\usepackage{color} 
\usepackage{tabu} 
\usepackage{longtable}
\usepackage{nomencl} 
\makenomenclature 
\biboptions{sort&compress}
\soulregister\citep7 
\soulregister\citet7 
\soulregister\citealp7 
\newsavebox{\measurebox} 
\newcommand{\vect}[1]{\boldsymbol{\mathbf{#1}}} 

\journal{Engineering Fracture Mechanics}

\makeatletter
\def\@author#1{\g@addto@macro\elsauthors{\normalsize%
    \def\baselinestretch{1}%
    \upshape\authorsep#1\unskip\textsuperscript{%
      \ifx\@fnmark\@empty\else\unskip\sep\@fnmark\let\sep=,\fi
      \ifx\@corref\@empty\else\unskip\sep\@corref\let\sep=,\fi
      }%
    \def\authorsep{\unskip,\space}%
    \global\let\@fnmark\@empty
    \global\let\@corref\@empty  
    \global\let\sep\@empty}%
    \@eadauthor={#1}
}
\makeatother

\begin{document}

\begin{frontmatter}



\title{A cohesive zone framework for environmentally assisted fatigue}


\author{Susana {del Busto}\fnref{Uniovi}}

\author{Covadonga Beteg\'on\fnref{Uniovi}}

\author{Emilio Mart\'{\i}nez-Pa\~neda\corref{cor1}\fnref{DTU}}
\ead{mail@empaneda.com}

\address[Uniovi]{Department of Construction and Manufacturing Engineering, University of Oviedo, Gij\'on 33203, Spain}

\address[DTU]{Department of Mechanical Engineering, Technical University of Denmark, DK-2800 Kgs. Lyngby, Denmark}

\cortext[cor1]{Corresponding author. Tel: +45 45 25 42 71; fax: +45 25 19 61.}

\begin{abstract}
We present a compelling finite element framework to model hydrogen assisted fatigue by means of a hydrogen- and cycle-dependent cohesive zone formulation. The model builds upon: (i) appropriate environmental boundary conditions, (ii) a coupled mechanical and hydrogen diffusion response, driven by chemical potential gradients, (iii) a mechanical behavior characterized by finite deformation J2 plasticity, (iv) a phenomenological trapping model, (v) an irreversible cohesive zone formulation for fatigue, grounded on continuum damage mechanics, and (vi) a traction-separation law dependent on hydrogen coverage calculated from first principles. The computations show that the present scheme appropriately captures the main experimental trends; namely, the sensitivity of fatigue crack growth rates to the loading frequency and the environment. The role of yield strength, work hardening, and constraint conditions in enhancing crack growth rates as a function of the frequency is thoroughly investigated. The results reveal the need to incorporate additional sources of stress elevation, such as gradient-enhanced dislocation hardening, to attain a quantitative agreement with the experiments.
\end{abstract}


\end{frontmatter}



\begin{framed}
\nomenclature{$a$}{crack length}
\nomenclature{$b, \, b_0$}{current and initial crack opening displacement}
\nomenclature{$\vect{B}$}{standard strain-displacement matrix}
\nomenclature{$\vect{B_c}$}{global cohesive displacement-separation matrix}
\nomenclature{$C$}{total hydrogen concentration}
\nomenclature{$C_L, \, C_T$}{hydrogen concentration in lattice and trapping sites}
\nomenclature{$c_q$}{specific heat capacity}
\nomenclature{$\mathcal{C}, \, m$}{Paris law coefficients}
\nomenclature{$\mathcal{D}, \, \mathcal{D}_e$}{standard and effective diffusion coefficients}
\nomenclature{$D, \, D_c, \, D_m$}{damage variable: total, cyclic and monotonic}
\nomenclature{$E$}{Young's modulus}
\nomenclature{$f$}{load frequency}
\nomenclature{$\vect{f_c}$}{cohesive internal force vector}
\nomenclature{$\Delta g_b^0$}{Gibbs free energy difference}
\nomenclature{$\vect{J}$}{hydrogen flux vector}
\nomenclature{$\vect{K_c}$}{cohesive tangent stiffness matrix}
\nomenclature{$K_T$}{trap equilibrium constant}
\nomenclature{$K, \, K_0$}{remote and reference stress intensity factor}
\nomenclature{$\vect{L}$}{local displacement-separation matrix}
\nomenclature{$\vect{\mathscr{L}}$}{elastoplastic constitutive matrix}
\nomenclature{$\vect{N}$}{shape functions matrix}
\nomenclature{$N$}{number of cycles}
\nomenclature{$\mathcal{N}$}{strain hardening exponent}
\nomenclature{$N_A$}{Avogadro's number}
\nomenclature{$N_L, \, N_T$}{number of lattice and trapping sites per unit volume}
\nomenclature{$q$}{heat flux per unit area}
\nomenclature{$\mathcal{R}$}{universal gas constant}
\nomenclature{$\vect{R}$}{rotational matrix}
\nomenclature{$R$}{load ratio}
\nomenclature{$R_0$}{reference plastic length}
\nomenclature{$T$}{elastic T-stress}
\nomenclature{$\mathcal{T}$}{absolute temperature}
\nomenclature{$\vect{T}, \, \tilde{\vect{T}}$}{standard and effective cohesive traction vectors}
\nomenclature{$\vect{t}$}{external traction vector}
\nomenclature{$T_n$}{normal cohesive traction}
\nomenclature{$U$}{internal energy per unit mass}
\nomenclature{$\vect{U}$}{global nodal displacement vector}
\nomenclature{$\vect{u}, \, \tilde{\vect{u}}$}{field and local nodal displacement vectors}
\nomenclature{$\bar{V}_H$}{partial molar volume of hydrogen}
\nomenclature{$V_M$}{molar volume of the host lattice}
\nomenclature{$W_B$}{trap binding energy}
\nomenclature{$\alpha$}{compression penalty factor}
\nomenclature{$\beta$}{number of lattice sites per solvent atom}
\nomenclature{$\vect{\Delta}, \, \tilde{\vect{\Delta}}$}{local field and nodal separation vectors}
\nomenclature{$\Delta_n$}{normal cohesive separation}
\nomenclature{$\delta_n$}{characteristic normal cohesive length}
\nomenclature{$\delta_{\Sigma}$}{accumulated cohesive length}
\nomenclature{$\varepsilon_p$}{equivalent plastic strain}
\nomenclature{$\vect{\varepsilon}$}{Cauchy strain tensor}
\nomenclature{$\theta_H$}{hydrogen coverage}
\nomenclature{$\theta_L, \, \theta_T$}{occupancy of lattice and trapping sites}
\nomenclature{$\mu_L$}{lattice chemical potential}
\nomenclature{$\rho$}{density}
\nomenclature{$\sigma_f$}{cohesive endurance limit}
\nomenclature{$\sigma_H$}{hydrostatic stress}
\nomenclature{$\sigma_Y$}{initial yield stress}
\nomenclature{$\vect{\sigma}$}{Cauchy stress tensor}
\nomenclature{$\sigma_{max}, \, \sigma_{max,0}$}{current and original cohesive strength}
\nomenclature{$\phi_n$}{normal cohesive energy}
\printnomenclature
\end{framed}

\section{Introduction}
\label{Sec:Introduction}

Metallic materials play a predominant role in structures and industrial components because of their strength, stiffness, toughness and tolerance of high temperatures. However, hydrogen has been known for over a hundred years to severely degrade the fracture resistance of advanced alloys, with cracking being observed in modern steels at one-tenth of the expected fracture toughness. With current engineering approaches being mainly empirical and highly conservative, there is a strong need to understand the mechanisms of such hydrogen-induced degradation and to develop mechanistic-based models able to reproduce the microstructure-dependent mechanical response at scales relevant to engineering practice.\\

Models based on the hydrogen enhanced decohesion (HEDE) mechanism have proven to capture the main experimental trends depicted by high-strength steels in aqueous solutions and hydrogen-containing gaseous environments \cite{G03}. The use of cohesive zone formulations is particularly appealing in this regard, as they constitute a suitable tool to characterize the sensitivity of the fracture energy to hydrogen coverage. The cohesive traction separation law can be derived from first principles quantum mechanics \cite{S04} or calibrated with experiments \cite{S08,Y16}. The statistical distribution of relevant microstructural features has also fostered the use of weakest-link approaches \cite{N10,A14}. Very recently, Mart\'{\i}nez-Pa\~neda \textit{et al.} \cite{M16} integrated strain gradient plasticity simulations and electrochemical assessment of hydrogen solubility in Gerberich \cite{G12} model. The investigation of a Ni-Cu superalloy and a modern ultra-high-strength steel revealed an encouraging quantitative agreement with experimental data for the threshold stress intensity factor and the stage II crack growth rate. However, and despite the fact that most industrial components experience periodic loading, modeling efforts have been mainly restricted to monotonic conditions. Recently, Moriconi \textit{et al.} \cite{M14} conducted experiments and simulations to investigate the role of hydrogen on a 15-5PH martensitic steel intended for gaseous hydrogen storage. Model predictions provided a very good agreement with experimental data for low hydrogen pressures but failed to capture the deleterious effect of hydrogen on the fatigue crack propagation under high pressures. Understanding the role of hydrogen in accelerating crack growth rates under cyclic loading could be crucial to enable the use of high-strength steels in the energy sector and to develop reliable transport and storage infrastructure for future energy systems.\\

In this work, we present a general numerical framework for hydrogen-assisted fatigue. The main ingredients of the model are: (i) realistic Dirichlet type conditions to account for stress-assisted diffusion at the boundaries, (ii)  an extended hydrogen transport equation governed by hydrostatic stresses and plastic straining through trapping, (iii) higher order elements incorporating a coupled mechanical-diffusion response, (iv) continuum large strains elastoplasticity, (v) a hydrogen coverage dependent cohesive strength, and (vi) a Lemaitre-type damage response for an irreversible traction-separation law. The influence of diffusible hydrogen in fatigue crack growth is systematically investigated, the main experimental trends captured and valuable insight achieved.

\section{Numerical framework}
\label{Sec:NumModel}

Hydrogen transport towards the fracture process zone and subsequent cracking under cyclic loading conditions are investigated by means of a coupled mechanical-diffusion-cohesive finite element framework. Section \ref{Sec:UMATHT} describes the mechanical-diffusion coupling that builds upon the analogy with heat transfer, Section \ref{Sec:CZM} provides details of the cyclic and hydrogen dependent cohesive zone formulation employed and finally Section \ref{Sec:FEM} outlines the general assemblage and implementation. 

\subsection{Coupled mechanical-diffusion through the analogy with heat transfer}
\label{Sec:UMATHT}

The hydrogen transport model follows the pioneering work by Sofronis and McMeeking \cite{SM89}. Hence, hydrogen transport is governed by hydrostatic stress and plastic straining through trapping. Hydrogen moves through normal interstitial lattice site diffusion and the diffusible concentration of hydrogen $C$ is defined as the sum of the hydrogen concentrations at reversible traps $C_T$ and lattice sites $C_L$. The latter is given by,
\begin{equation}\label{Eq:CL}
C_L=N_L \theta_L
\end{equation}

\noindent where $N_L$ is the number of sites per unit volume and $\theta_L$ the occupancy of lattice sites. The former can be expressed as a function of $V_M$, the molar volume of the host lattice, as:
\begin{equation}\label{Eq:NLVA}
N_L=\frac{\beta N_A}{V_M}
\end{equation} 

\noindent with $N_A$ being Avogadro's number and $\beta$ the number of interstitial lattice sites per solvent atom. On the other hand, the hydrogen concentration trapped at microstructural defects is given by,
\begin{equation}\label{Eq:CT}
C_T=N_T \theta_T 
\end{equation}

\noindent where $N_T$ denotes the number of traps per unit volume and $\theta_T$ the occupancy of the trap sites. Here, focus will be placed on reversible trapping sites at microstructural defects generated by plastic straining - dislocations; a key ingredient in the mechanics of hydrogen diffusion \cite{O14,D15}. A phenomenological relation between the trap density and the equivalent plastic strain is established based on the permeation tests by Kumnick and Johnson \cite{KJ80},
\begin{equation}\label{Eq:NTKJ}
\log N_T = 23.26 - 2.33 \exp \left( -5.5 \varepsilon_p \right)
\end{equation}   

Oriani's equilibrium theory \cite{O70} is adopted, resulting in a Fermi-Dirac relation between the occupancy of trap and lattice sites,
\begin{equation}\label{Eq:Oriani}
\frac{\theta_T}{1-\theta_T}=\frac{\theta_L}{1-\theta_L} K_T 
\end{equation}

\noindent with $K_T$ being the trap equilibrium constant,
\begin{equation}
K_T=\textnormal{exp}\left( \frac{-W_B}{\mathcal{R}\mathcal{T}} \right)
\end{equation}

Here, $W_B$ is the trap binding energy, $\mathcal{R}$ the gas constant and $\mathcal{T}$ the absolute temperature. Under the common assumption of low occupancy conditions ($\theta_L << 1$), the equilibrium relationship between $C_T$ and $C_L$ becomes,
\begin{equation}\label{Eq:CT2}
C_T=\frac{N_T K_T C_L}{K_T C_L + N_L}
\end{equation}

In a volume, $V$, bounded by a surface, $S$, with outward normal, $\vect{n}$, mass conservation requirements relate the rate of change of $C$ with the hydrogen flux through $S$,
\begin{equation}\label{Eq:MassBal}
\frac{\textnormal{d}}{\textnormal{d}t} \int_V C \,\, \textnormal{d}V + \int_S \vect{J} \cdot \vect{n} \,\, \textnormal{d} S=0
\end{equation}

Fick's law relates the hydrogen flux with the gradient of the chemical potential $\nabla \mu_L$,
\begin{equation}\label{Eq:J1}
\vect{J} = - \frac{\mathcal{D} C_L}{\mathcal{R}\mathcal{T}} \nabla \mu_L
\end{equation}

\noindent with $\mathcal{D}$ being the diffusion coefficient. The chemical potential of hydrogen in lattice sites is given by,
\begin{equation}\label{Eq:muL}
\mu_L = \mu_L^0 + \mathcal{R}\mathcal{T} \, \textnormal{ln} \frac{\theta_L}{1-\theta_L} - \bar{V}_H \sigma_H
\end{equation}

Here, $\mu_L^0$ denotes the chemical potential in the standard state and the last term corresponds to the so-called stress-dependent part of the chemical potential $\mu_{\sigma}$, with $\bar{V}_H$ being the partial molar volume of hydrogen in solid solution. Assuming a constant interstitial sites concentration and substituting (\ref{Eq:muL}) into (\ref{Eq:J1}), one reaches
\begin{equation}\label{Eq:JL2}
\vect{J}=-\mathcal{D} \nabla C_L + \frac{\mathcal{D}}{\mathcal{R}\mathcal{T}} C_L \bar{V}_H \nabla \sigma_H
\end{equation}

Replacing $\vect{J}$ in the mass balance equation (\ref{Eq:MassBal}), using the divergence theorem and considering the arbitrariness of $V$ renders,
\begin{equation}\label{Eq:Bal2}
\frac{dC_L}{dt}+\frac{dC_T}{dt} = \mathcal{D} \nabla^2 C_L  - \nabla \cdot \left( \frac{\mathcal{D} C_L \bar{V}_H}{\mathcal{R}\mathcal{T}}  \nabla \sigma_H \right)
\end{equation}

It is possible to phrase the left-hand side of (\ref{Eq:Bal2}) in terms of $C_L$ by making use of Oriani's theory of equilibrium,
\begin{equation}
\frac{\mathcal{D}}{\mathcal{D}_e} \frac{d C_L}{dt}= \mathcal{D} \nabla^2 C_L - \nabla \cdot \left( \frac{\mathcal{D} C_L \bar{V}_H}{\mathcal{R}\mathcal{T}}  \nabla \sigma_H \right)
\end{equation}

\noindent where an effective diffusion constant has been defined,
\begin{equation}\label{Eq:Deff}
\mathcal{D}_e=\mathcal{D} \frac{C_L}{C_L + C_T (1 - \theta_T)}
\end{equation}

Regarding the boundary conditions, a constant hydrogen concentration $C_b$ is prescribed at the crack faces in the vast majority of hydrogen embrittlement studies. However, as noted by Turnbull \cite{T15}, such scheme may oversimplify the electrochemistry-diffusion interface and the use of generalized boundary conditions is particularly recommended for materials with high hydrogen diffusivity. Here, we follow Mart\'{\i}nez-Pa\~neda \textit{et al.} \cite{M16b} and adopt Dirichlet-type boundary conditions where the lattice hydrogen concentration at the crack faces depends on the hydrostatic stress. Hence, the lattice hydrogen concentration at the crack faces equals,
\begin{equation}\label{eq:DISP}
C_L=C_b \exp \left( \frac{\bar{V}_H \sigma_H}{\mathcal{R}\mathcal{T}} \right)
\end{equation}

\noindent which is equivalent to prescribing a constant chemical potential. To this end, a user subroutine DISP is employed in ABAQUS to relate the magnitude of $C_L$ to a nodal averaged value of the hydrostatic stress. Also, the domain where the boundary conditions are enforced changes with crack advance. Consequently, a multi-point constraint (MPC) subroutine is defined to update the boundary region throughout the analysis - see Section \ref{Sec:FEM}.\\

Finite deformation J2 plasticity theory is used to compute the two mechanical ingredients of the present hydrogen transport scheme, $\varepsilon_p$ and $\sigma_H$. We develop a fully coupled mass transport - continuum elastoplastic finite element framework that is solved in a monolithic way. Higher order elements are used, with nodal displacements and lattice hydrogen concentration being the primary variables. The numerical implementation is carried out in the well-known finite element package ABAQUS. To this end, a UMATHT subroutine is developed to exploit the analogy with heat transfer \cite{B16,D16}. Thus, the energy balance for a stationary solid in the absence of heat sources is given by,
\begin{equation}
\int_V \rho \, \dot{U} \, \textnormal{d}V - \int_S q \, \textnormal{d}S = 0
\end{equation}

\noindent where $\rho$ is the density, $q$ the heat flux per unit area of the solid and $\dot{U}$ the material time rate of the internal energy, the latter being related to the temperature change through the specific heat capacity: $\dot{U}=c_q \dot{\mathcal{T}}$. The similitude with (\ref{Eq:MassBal}) is clear and an appropriate analogy can be easily established (see Table \ref{Tab:DiffusionHeat}), enabling the use of the coupled temperature-displacement capabilities already available in ABAQUS.

\begin{table}[H]
\centering
\caption{Analogy between heat transfer and mass diffusion.}
\label{Tab:DiffusionHeat}
   {\tabulinesep=1.2mm
   \begin{tabu} {cc}
       \hline
 Heat transfer & Mass diffusion \\ \hline
 $\rho c_p \frac{\partial \mathcal{T}}{\partial t} + \nabla q = 0$  &   $\frac{\partial C_i}{\partial t} + \nabla \vect{J}=0$\\
 $\dot{U}=c_p \dot{\mathcal{T}}$ & $\frac{\partial C}{\partial t}=\frac{\partial (C_L + C_T)}{\partial t}$  \\
 $\mathcal{T}$ & $C_L$  \\
 $c_p$ & $\mathcal{D}/\mathcal{D}_{e}$ \\
 $\rho$ & 1  \\\hline
   \end{tabu}}
\end{table}

\subsection{Cohesive zone model}
\label{Sec:CZM}

A cohesive zone formulation will be employed to model crack initiation and subsequent growth. Based on the pioneering works by Dugdale \cite{D60} and Barenblatt \cite{B62}, cohesive zone models introduce the notion of a cohesive force ahead of the crack that prevents propagation. The micromechanisms of material degradation and failure are thus embedded into the constitutive law that relates the cohesive traction with the local separation. Damage is restricted to evolve along the pre-defined cohesive interface, and consequently, the numerical implementation is generally conducted by inserting cohesive finite elements in potential crack propagation paths. Hence, in the absence of body forces, the weak form of the equilibrium equations for a body of volume $V$ and external surface $S$ renders,
\begin{equation}\label{Eq:CoheWeak}
\int_V \vect{\sigma} : \delta \vect{\varepsilon}  \, \textnormal{d}V + \int_{S_c} \vect{T} \cdot \delta \vect{\Delta} \, \textnormal{d}S= \int_S \vect{t} \cdot \delta \vect{u} \, \textnormal{d}S
\end{equation}

Here, $\vect{T}$ are the cohesive tractions and $S_c$ is the surface across which these tractions operate. The standard part of the mechanical equilibrium statement is characterized by the Cauchy stress tensor $\vect{\sigma}$, the work-conjugate strain tensor $\vect{\varepsilon}$, the external tractions $\vect{t}$ and the displacement vector $\vect{u}$; the latter being obtained by interpolating the global nodal displacement $\vect{u}=\vect{N}\vect{U}$. The local nodal separation $\tilde{\vect{\Delta}}$ is related to the local nodal displacement $\tilde{\vect{u}}$ by
\begin{equation}
\tilde{\vect{\Delta}} = \vect{L} \tilde{\vect{u}}
\end{equation}

\noindent where $\vect{L}$ is a local displacement-separation relation matrix. The separation along a cohesive surface element is interpolated from the nodal separation by means of standard shape functions,
\begin{equation}
\vect{\Delta} = \vect{N} \tilde{\vect{\Delta}}
\end{equation}

\noindent and the global nodal displacement is related to the local nodal displacement by means of a rotational matrix:
\begin{equation}
\tilde{\vect{u}} = \vect{R} \vect{U}
\end{equation}

The relationship between the local separation and the global nodal displacement can be then obtained by combining the previous equations,
\begin{equation}
\vect{\Delta} = \vect{B}_c \vect{U}
\end{equation}

\noindent where $\vect{B}_c$ is a global displacement-separation relation matrix: $\vect{B}_c=\vect{N} \vect{R} \vect{L}$. Thus, accounting for the classic finite element discretization in (\ref{Eq:CoheWeak}) and requiring the variational statement to hold for any admissible field, it renders
\begin{equation}
\int_V \vect{B}^T \vect{\mathscr{L}} \vect{\varepsilon}  \, \textnormal{d}V + \int_{S_c} \vect{B}_c^T \vect{T} \, \textnormal{d}S= \int_S \vect{N}^T \vect{t} \, \textnormal{d}S
\end{equation}

\noindent where $\vect{\mathscr{L}}$ is the elastoplastic constitutive matrix and $\vect{B}$ the standard strain-displacement matrix. Considering the dependence of $\vect{\varepsilon}$ and $\vect{T}$ on $\vect{U}$,
\begin{equation}
\vect{U} \left( \int_V \vect{B}^T \vect{\mathscr{L}} \vect{B} \, \textnormal{d}V + \int_{S_c} \vect{B}_c^T \frac{\partial \vect{T}}{\partial \vect{\Delta}} \vect{B}_c \, \textnormal{d}S \right) = \int_S \vect{N}^T \vect{t} \, \textnormal{d}S
\end{equation}

\noindent and the components of the classic finite element global system of equations can be readily identified. The stiffness matrix of the cohesive elements is therefore given by,
\begin{equation}
\vect{K}_c = \int_{S_c} \vect{B}_c^T \frac{\partial \vect{T}}{\partial \vect{\Delta}} \vect{B}_c \, \textnormal{d}S
\end{equation}

\noindent which corresponds to the gradient of the internal cohesive force vector,
\begin{equation}
\vect{f}_c=\int_{S_0} \vect{B}_c^T \vect{T} dS
\end{equation} 

\begin{figure}[H]
\centering
\includegraphics[scale=0.8]{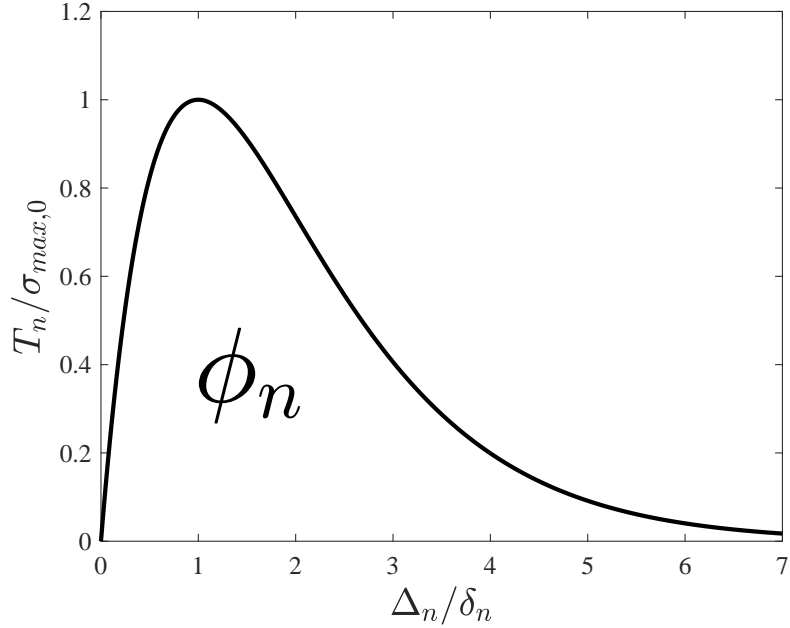}
\caption{Traction-separation law characterizing the cohesive zone model in the absence of cyclic loading and hydrogen degradation.}
\label{fig:CoheLaw}
\end{figure}

The pivotal ingredient of cohesive zone models is the traction-separation law that governs material degradation and separation. The exponentially decaying cohesive law proposed by Xu and Needleman \cite{XN93} is here adopted. Focus will be placed on pure mode I problems and consequently, the constitutive equations related to the tangential separation will be omitted for the sake of brevity. As depicted in Fig. \ref{fig:CoheLaw}, for a given shape of the traction-separation curve, the cohesive response can be fully characterized by two parameters, the cohesive energy $\phi_n$ and the critical cohesive strength $\sigma_{max,0}$.\\

The cohesive response is therefore characterized by the relation between the normal tractions ($T_n$) and the corresponding displacement jump ($\Delta_n$) as,
\begin{equation}
T_n=\frac{\phi_n}{\delta_n} \left( \frac{\Delta_n}{\delta_n} \right) \exp \left( - \frac{\Delta_n}{\delta_n} \right)
\end{equation}

\noindent with the normal work of separation $\phi_n$ being given by,
\begin{equation}
\phi_n= \exp(1) \sigma_{max,0} \delta_n
\end{equation}

\noindent where $\delta_N$ is the characteristic cohesive length under normal separation. The effect of hydrogen in lowering the cohesive strength, and subsequently the fracture toughness, is captured here by employing the impurity-dependent cohesive law proposed by Serebrinsky et al. \cite{S04}. Hence, a first-principles-based relation between the \emph{current} cohesive strength $\sigma_{max}$ and the original cohesive strength in the absence of hydrogen $\sigma_{max,0}$ is defined,
\begin{equation}
\frac{\sigma_{max}(\theta_H)}{\sigma_{max,0}}=1 - 1.0467 \theta_H + 0.1687 \theta_H^2
\end{equation}

\noindent where $\theta_H$ is the hydrogen coverage, which is defined as a function of hydrogen concentration and Gibbs free energy difference between the interface and the surrounding material, as expressed in the Langmuir-McLean isotherm:
\begin{equation}
\theta_H=\frac{C}{C+\exp \left( \frac{- \Delta g_b^0}{\mathcal{R}\mathcal{T}} \right)}
\end{equation}

\begin{figure}[H]
\centering
\includegraphics[scale=0.9]{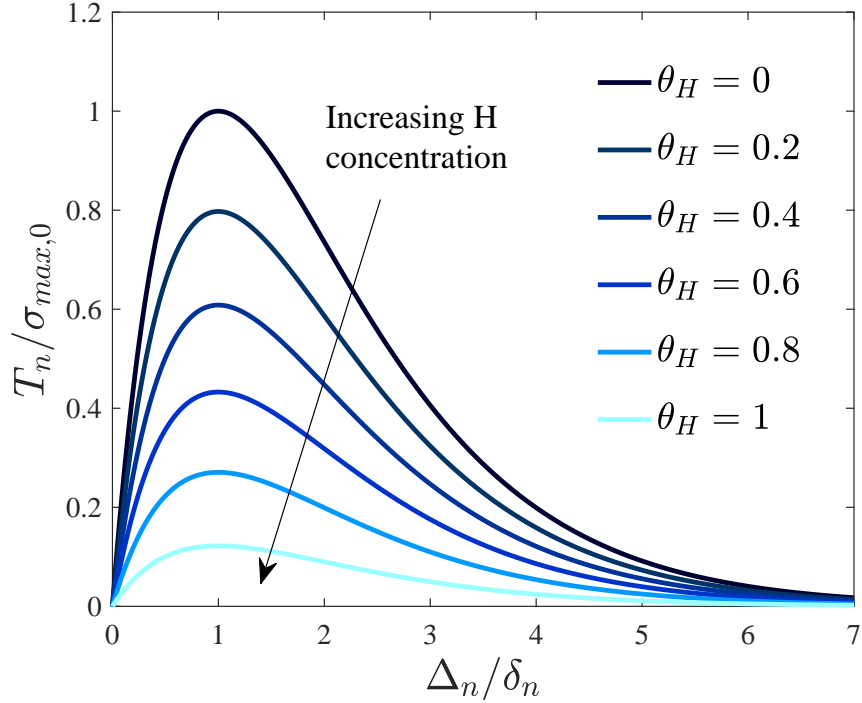}
\caption{Effect of hydrogen coverage $\theta_H$ on the traction-separation law characterizing the cohesive response.}
\label{fig:CoheLawH}
\end{figure}

A value of 30 kJ/mol is assigned to the trapping energy $\Delta g_b^0$ in \citep{S04} from the spectrum of experimental data available. Thus, from first principles calculations of hydrogen atoms in bcc Fe, a quantum-mechanically informed traction-separation law can be defined as a function of the hydrogen coverage \cite{S04} (see Fig. \ref{fig:CoheLawH}).\\ 

On the other hand, cyclic damage is incorporated by means of the irreversible cohesive zone model proposed by Roe and Siegmund \cite{RS03}. The model incorporates (i) loading-unloading conditions, (ii) accumulation of damage during subcritical cyclic loading, and (iii) crack surface contact. A damage mechanics approach is adopted to capture the cohesive properties degradation as a function of the number of cycles. A damage variable $D$ is defined so that it represents the effective surface density of micro defects in the interface. Consequently, an effective cohesive zone traction can be formulated: $\tilde{\vect{T}}=\vect{T}/(1-D)$. Subsequently, the current or effective cohesive strength $\sigma_{max}$ is related to the initial cohesive strength $\sigma_{max,0}$ as,
\begin{equation}
\sigma_{max}=\sigma_{max,0} (1 - D)
\end{equation}

\begin{figure}[H]
\centering
\includegraphics[scale=0.9]{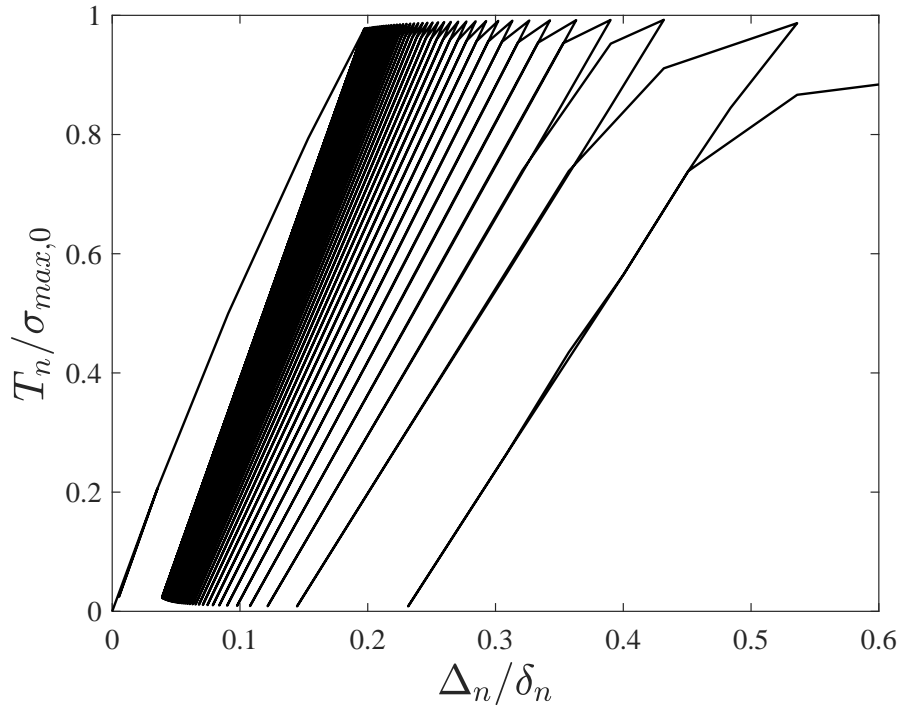}
\caption{Cohesive response under stress-controlled cyclic loading conditions.}
\label{fig:CoheLawF}
\end{figure}

A damage evolution law is defined so that it incorporates the relevant features of continuum damage approaches, namely: (i) damage accumulation starts if a deformation measure is greater than a critical magnitude, (ii) the increment of damage is related to the increment of deformation, and (iii) an endurance limit exists, bellow which cyclic loading can proceed infinitely without failure. From these considerations, cyclic damage evolution is defined as,
\begin{equation}\label{Eq:Damage}
\dot{D}_c= \frac{|\dot{\Delta}_n|}{\delta_{\Sigma}} \left[ \frac{T_n}{\sigma_{max}} - \frac{\sigma_f}{\sigma_{max,0}} \right] H \left( \bar{\Delta}_n -  \delta_n \right)
\end{equation}

\noindent with $\bar{\Delta}_n=\int |\dot{\Delta}_n| dt$ and $H$ denoting the Heaviside function. Two new parameters have been introduced: $\sigma_f$, the cohesive endurance limit and $\delta_{\Sigma}$, the accumulated cohesive length - used to scale the normalized increment of the effective material separation. The modeling framework must also incorporate damage due to monotonic loading; as a consequence, the damage state is defined as the maximum of the cyclic and monotonic contributions,
\begin{equation}
D= \int \textnormal{max} \left( \dot{D}_c, \dot{D}_m \right) dt
\end{equation}

\noindent being the latter characterized as:
\begin{equation}
\dot{D}_m = \frac{ \left. \textnormal{max} \left( \Delta_n \right) \right|_{t_i} -  \left. \textnormal{max} \left( \Delta_n \right) \right|_{t_{i-1}}}{4 \delta_n}
\end{equation}

\noindent and updated only when the largest stored value of $\Delta_n$ is greater than $\delta_N$. Here, $t_{i-1}$ denotes the previous time increment and $t_i$ the current one. In addition to damage evolution, the cohesive response must be defined for the cases of unloading/reloading, compression, and contact between the crack faces. Unloading is defined based on the analogy with an elastic-plastic material undergoing damage. Thereby, unloading takes place with the stiffness of the cohesive zone at zero separation, such that
\begin{equation}
T_n=T_{max} + \left( \frac{\exp(1) \sigma_{max}}{\delta_n} \right) \left( \Delta_n - \Delta_{max} \right)
\end{equation}

\noindent where $\Delta_{max}$ is the maximum separation value that has been attained and $T_{max}$ its associated traction quantity. Compression behavior applies when the unloading path reaches $\Delta_n=0$ at $T_n < 0$. In such circumstances, the cohesive response is given by,
\begin{align}
T_n = & \frac{\phi_n}{\delta_n} \left( \frac{\Delta_n}{\delta_n} \right) \exp \left( - \frac{\Delta_n}{\delta_n} \right) + T_{max} - \sigma_{max} \exp(1) \frac{\Delta_{max}}{\delta_n} \nonumber \\
& + \alpha \sigma_{max,0} \exp(1) \frac{\Delta_n}{\delta_n} \exp \left( - \frac{\Delta_n}{\delta_n} \right)
\end{align}

\noindent being $\alpha$ a penalty factor that is taken to be equal to 10, following \cite{RS03}. Contact conditions are enforced if $\Delta_n$ is negative and the cohesive element has failed completely ($D=1$). At this instance the cohesive law renders,
\begin{equation}
T_n= \alpha \sigma_{max,0} \exp(1) \exp \left( - \frac{\Delta_n}{\delta_n} \right) \frac{\Delta_n}{\delta_n} 
\end{equation}

\noindent where friction effects have been neglected. Fig. \ref{fig:CoheLawF} shows a representative response obtained by applying a stress-controlled cyclic loading $\Delta \sigma / \sigma_{max,0}=1$ with a zero stress ratio. The accumulated separation increases with the number of loading cycles, so that it becomes larger than $\delta_n$ and damage starts to play a role, lowering the stiffness and the cohesive strength.\\

This novel cyclic- and hydrogen concentration-dependent cohesive zone framework is implemented in ABAQUS by means of a user element UEL subroutine. The code can be downloaded from www.empaneda.com/codes and is expected to be helpful to both academic researchers and industry practitioners.\\

In some computations, numerical convergence is facilitated by employing the viscous regularization technique proposed by Gao and Bower \cite{GB04}. Such scheme leads to accurate results if the viscosity coefficient, $\xi$, is sufficiently small \cite{Y16}. A sensitivity study has been conducted in the few cases where viscous regularization was needed; values of $\xi$ on the order of $10^{-6}$ have proven to be appropriate for the boundary value problem under consideration. Other approaches to overcome snap-back instabilities, less suitable for cyclic loading, include the use of explicit finite element solution schemes \cite{S16} or determining the equilibrium path for a specified crack tip opening by means of control algorithms \cite{T76,SL04,M17b}.

\subsection{Finite element implementation}
\label{Sec:FEM}

\begin{figure}[H]
  \makebox[\textwidth][c]{\includegraphics[width=1.2\textwidth]{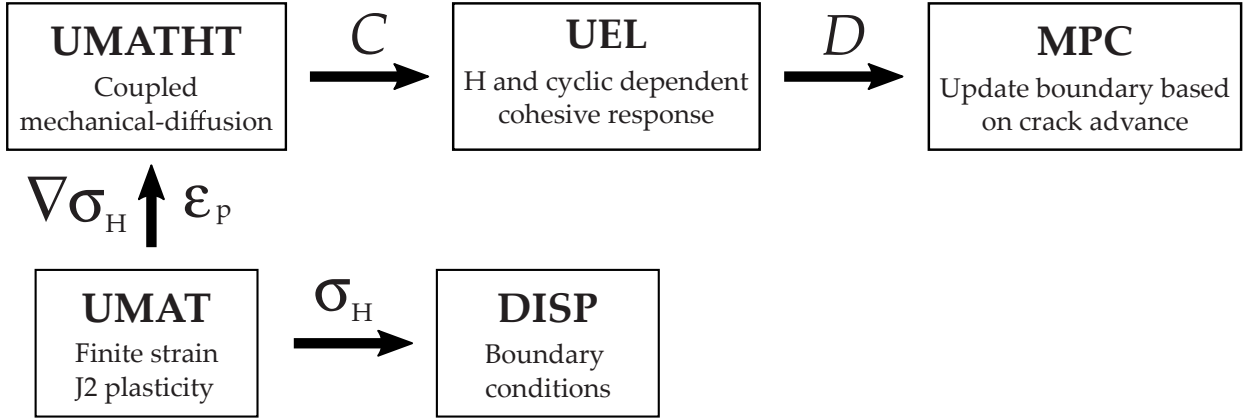}}%
  \caption{Schematic overview of the relations between the different Abaqus subroutines.}
  \label{fig:Abaqus}
\end{figure}

The aforementioned mechanical-diffusion-cohesive numerical framework is implemented in the commercial finite element package ABAQUS. Fortran modules are widely employed to transfer information between the different user subroutines. Thus, as described in Fig. \ref{fig:Abaqus}, a user material UMAT subroutine is developed to characterize the mechanical response by means of a finite strain version of conventional von Mises plasticity. The nodal averaged value of the hydrostatic stress at the crack faces is then provided to a DISP subroutine, so as to prescribe a more realistic $\sigma_H$-dependent lattice hydrogen concentration. The hydrostatic stress gradient is computed by means of linear shape functions and, together with the equivalent plastic strain, is afterward given as input to the UMATHT subroutine to capture the effects of chemical expansion and trapping. The UMATHT subroutine provides the cohesive elements with the diffusible concentration of hydrogen in their adjacent continuum element. The damage variable is then transferred from the user elements to the MPC subroutine to keep track of the crack extension. Multi-point constraints have been defined between the nodes ahead of the crack and a set of associated dummy nodes that are activated as the crack advances. Hydrogen diffusion is assumed to be instantaneous, such that the lattice hydrogen concentration at the boundary is immediately prescribed when a new portion of crack surface is available.\\

Higher order elements are used in all cases: 8-node quadrilateral elements with reduced integration are employed to model the bulk response, and crack initiation and growth are captured by 6-node quadrilateral cohesive elements with 12 integration points. Results post-processing is carried out in MATLAB by making use of \emph{Abaqus2Matlab} \cite{P17}, a novel tool that connects the two aforementioned well-known software suites. 

\section{Results}
\label{Sec:Results}

We investigate the pernicious effect of hydrogen in fatigue crack growth, of great relevance in both energy storage and transport. The synergistic interaction of cyclic plastic deformation and local hydrogen uptake is particularly detrimental, with catastrophic failure being observed in cases where hydrogen-assisted cracking is negligible under monotonic loading \cite{G90}.\\

\begin{figure}[H]
  \makebox[\textwidth][c]{\includegraphics[width=1.2\textwidth]{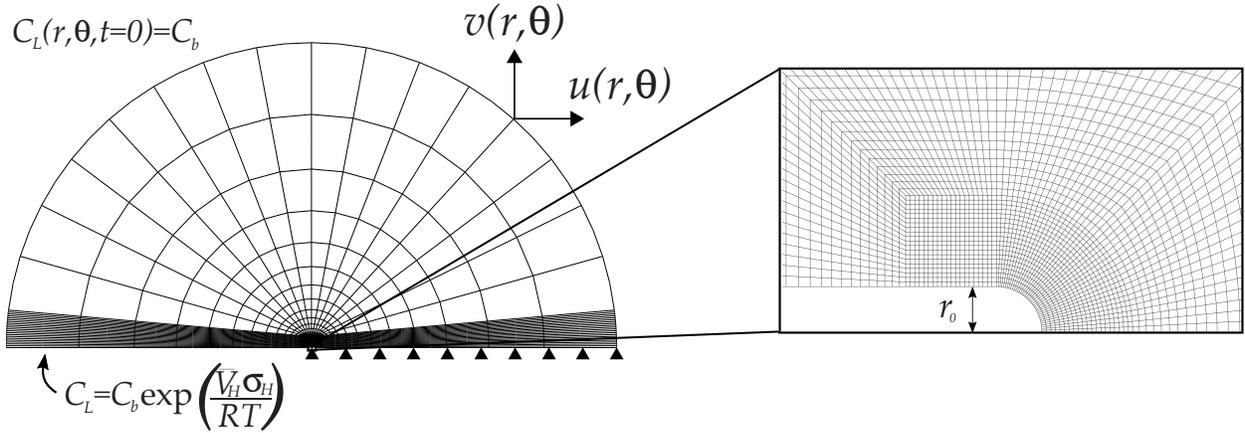}}%
  \caption{General and detailed representation of the finite element mesh employed for
the boundary layer model. Mechanical and diffusion boundary conditions are shown superimposed.}
  \label{fig:Mesh1}
\end{figure}

The boundary layer model employed by Sofronis and McMeeking \cite{SM89} is taken as a benchmark. Hence, hydrogen transport and subsequent cracking are investigated in an iron-based material with a diffusion coefficient of $\mathcal{D}=0.0127$ mm$^2$/s, Young's modulus of $E=207$ GPa, Poisson's ratio of $\nu=0.3$ and initial yield stress of $\sigma_Y=250$ MPa. Work hardening is captured by means of the following isotropic power law,
\begin{equation}
\sigma = \sigma_Y \left(1 + \frac{E \varepsilon_p}{\sigma_Y} \right)^\mathcal{N}
\end{equation}

\noindent with the strain hardening exponent being equal to $\mathcal{N}=0.2$. Isotropic hardening has been adopted to reproduce the conditions of \cite{SM89}, but one should note that other plastic flow models can be easily incorporated; the use of non-linear kinematic hardening laws is particularly convenient to appropriately capture the Bauschinger effect under low load ratios. As described in Fig. \ref{fig:Mesh1}, the crack region is contained within a circular zone and a remote Mode I load is applied by prescribing the displacements of the nodes at the outer boundary,
\begin{equation}
u \left( r, \theta \right) = K_I \frac{1+\nu}{E} \sqrt{\frac{r}{2 \pi}} \cos \left( \frac{\theta}{2} \right) \left(3 - 4 \nu - \cos \theta \right)
\end{equation}
\begin{equation}
v \left( r, \theta \right) = K_I \frac{1+\nu}{E} \sqrt{\frac{r}{2 \pi}} \sin \left( \frac{\theta}{2} \right) \left(3 - 4 \nu - \cos \theta \right)
\end{equation}

\noindent where $u$ and $v$ are the horizontal and vertical components of the displacement boundary condition, $r$ and $\theta$ the radial and angular coordinates of each boundary node in a polar coordinate system centered at the crack tip, and $K_I$ is the applied stress intensity factor that quantifies the remote load in small scale yielding conditions. The lattice hydrogen concentration is prescribed in the crack surface as a function of $\sigma_H$ and the boundary concentration in the absence of hydrostatic stresses $C_b$. Following \cite{SM89}, an initial bulk concentration equal to $C_b$ is also defined in the entire specimen at the beginning of the analysis. Only the upper half of the circular domain is modeled due to symmetry and the outer radius is chosen to be significantly larger than the initial crack tip blunting radius. As shown in Fig. \ref{fig:Mesh1}, a very refined mesh is used, with the characteristic element size in the vicinity of the crack, $l_e$, being significantly smaller than a reference plastic length,
\begin{equation}
R_0 = \frac{1}{3 \pi \left( 1 - \nu^2 \right)} \frac{E \phi_n}{\sigma_Y^2}
\end{equation}

\noindent ($l_e<2000R_0$). A sensitivity study is conducted to ensure that the mesh resolves the cohesive zone size - approximately 14000 quadrilateral 8-node plane strain elements are employed. The modeling framework is suitable for both low and high cycle fatigue, with computations of $10^4$ cycles (with at least 10 load increments per cycle) running overnight on a single core.\\

\begin{figure}[H]
        \centering
        \begin{subfigure}[h]{1\textwidth}
                \centering
                \includegraphics[scale=0.8]{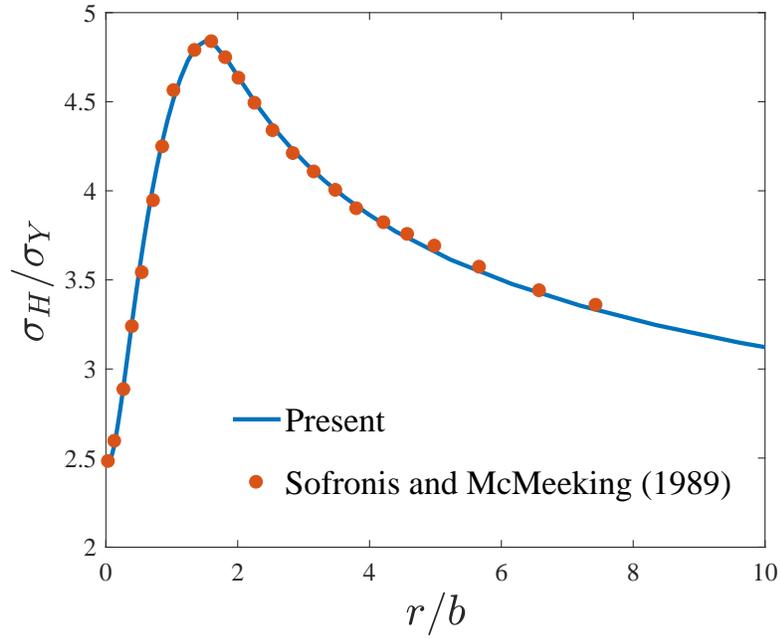}
                \caption{}
                \label{fig:SH}
        \end{subfigure}
        
        \begin{subfigure}[h]{1\textwidth}
                \centering
                \includegraphics[scale=0.8]{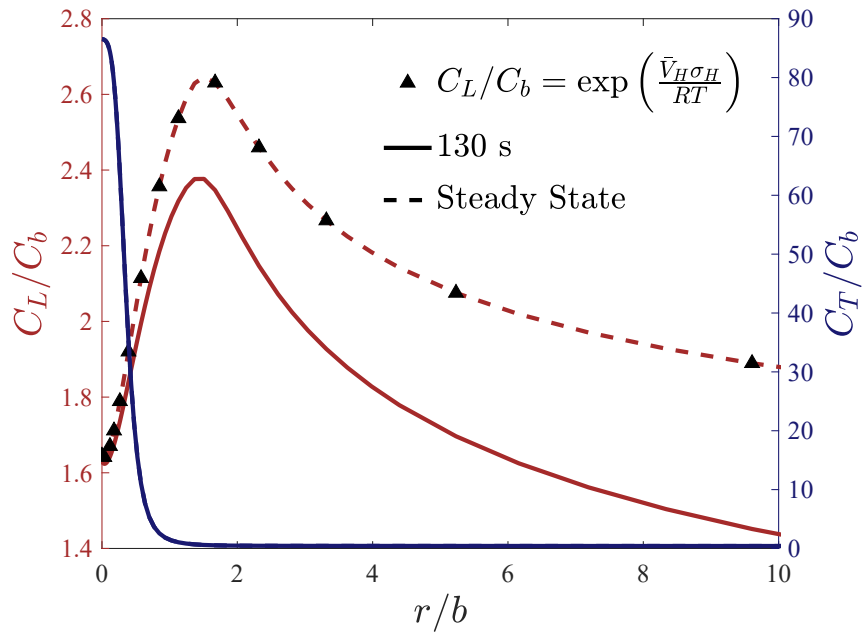}
                \caption{}
                \label{fig:CLCT}
        \end{subfigure}
       
        \caption{Crack tip fields for a stationary crack in an iron-based material under monotonic loading conditions, (a) normalized hydrostatic stress distribution for $K_I=2836.7$ MPa$\sqrt{mm}$ and (b) lattice and trap sites hydrogen concentrations at steady state and after 130 s.}\label{fig:CrackTipFields}
\end{figure}

We first validate the coupled mechanical-diffusion implementation by computing crack tip fields under monotonic loading conditions in the absence of crack propagation. Thus, the load is increased from zero at a rate of 21.82 MPa$\sqrt{mm} \,$s$^{-1}$ for 130 s and held fixed afterward, when the crack opening displacement is approximately 10 times the initial blunting radius $b=5b_0=10r_0$. Fig. \ref{fig:SH} shows the estimated hydrostatic stress distribution along with the predictions by Sofronis and McMeeking \cite{SM89} (symbols); results are shown along the extended crack plane with the distance to the crack tip normalized by the current crack tip opening $b$. A very good agreement is observed, verifying the finite strains J2 plasticity implementation. Fig. \ref{fig:CLCT} shows the results obtained for the lattice and trapped hydrogen concentrations for a boundary concentration of $C_b=2.08 \cdot 10^{12}$ H atoms/mm$^3$ at 130 s and after reaching steady-state conditions. The quantitative response described by the lattice hydrogen concentration when accounting for the dilatation of the lattice significantly differs to that obtained prescribing a constant $C_L$, as highlighted by Di Leo and Anand \cite{DA13} in the context of their constant lattice chemical potential implementation. The results achieved by means of the present $\sigma_H$-dependent Dirichlet scheme accurately follow the analytical steady-state solution for the distribution of the lattice hydrogen concentration ahead of the crack. On the other hand, $C_T$ shows a high peak at the crack tip and negligible sensitivity to the diffusion time (the curves for 130 s and steady state fall on top of each other); this is due to the governing role of plastic deformation as a result of the direct proportional relationship between $C_T$ and $N_T$.\\

Environmentally assisted fatigue is subsequently investigated by scaling in time the external load by a sinusoidal function. The cyclic boundary conditions prescribed are characterized by the load amplitude $\Delta K = K_{max} - K_{min}$ and the load ratio $R=K_{min}/K_{max}$. An initial prestressing is defined, such that the mean load equals the load amplitude, and both $R$ and $\Delta K$ remain constant through the analysis.\\

Following \cite{RS03}, the accumulated cohesive length in (\ref{Eq:Damage}) is chosen to be $\delta_{\Sigma}=4 \delta_n$ and the endurance coefficient $\sigma_f / \sigma_{max,0}=0.25$. The initial cohesive strength is assumed to be equal to $\sigma_{max,0}=3.5 \sigma_Y$ based on the seminal work by Tvergaard and Hutchinson \cite{TH92}. One should, however, note that such magnitude is intrinsically associated with the stress bounds of conventional plasticity - more realistic values can be obtained if the role of the increased dislocation density associated with large gradients in plastic strain near the crack tip is accounted for \cite{M17b,MB15}. A reference stress intensity factor,
\begin{equation}
K_0 = \sqrt{ \frac{E \phi_n}{(1-\nu^2)}}
\end{equation}

\noindent is defined to present the results.\\

\begin{figure}[H]
        \centering
        \begin{subfigure}[h]{1\textwidth}
                \centering
                \includegraphics[scale=0.8]{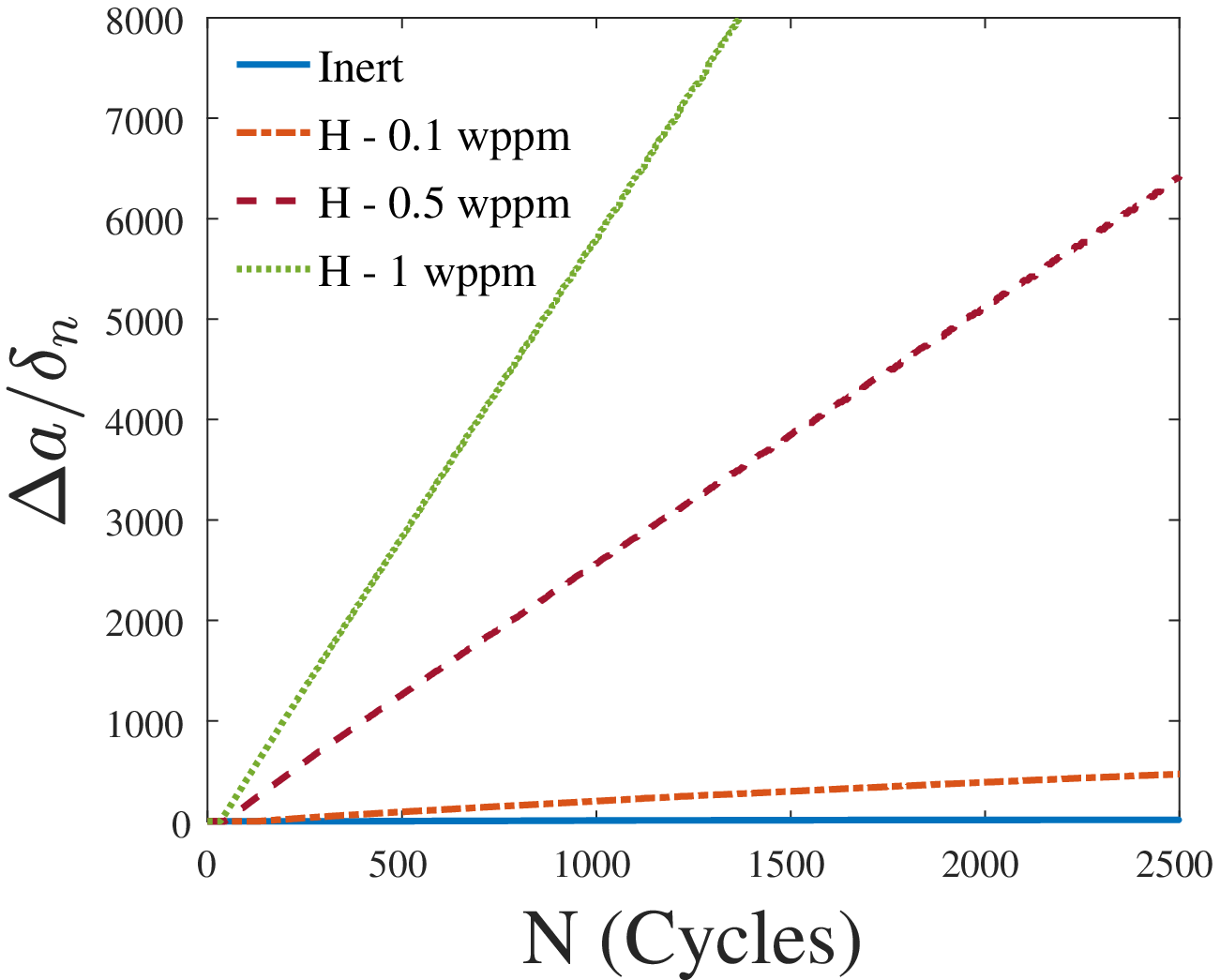}
                \caption{}
                \label{fig:InfluenceHa}
        \end{subfigure}
        
        \begin{subfigure}[h]{1\textwidth}
                \centering
                \includegraphics[scale=0.8]{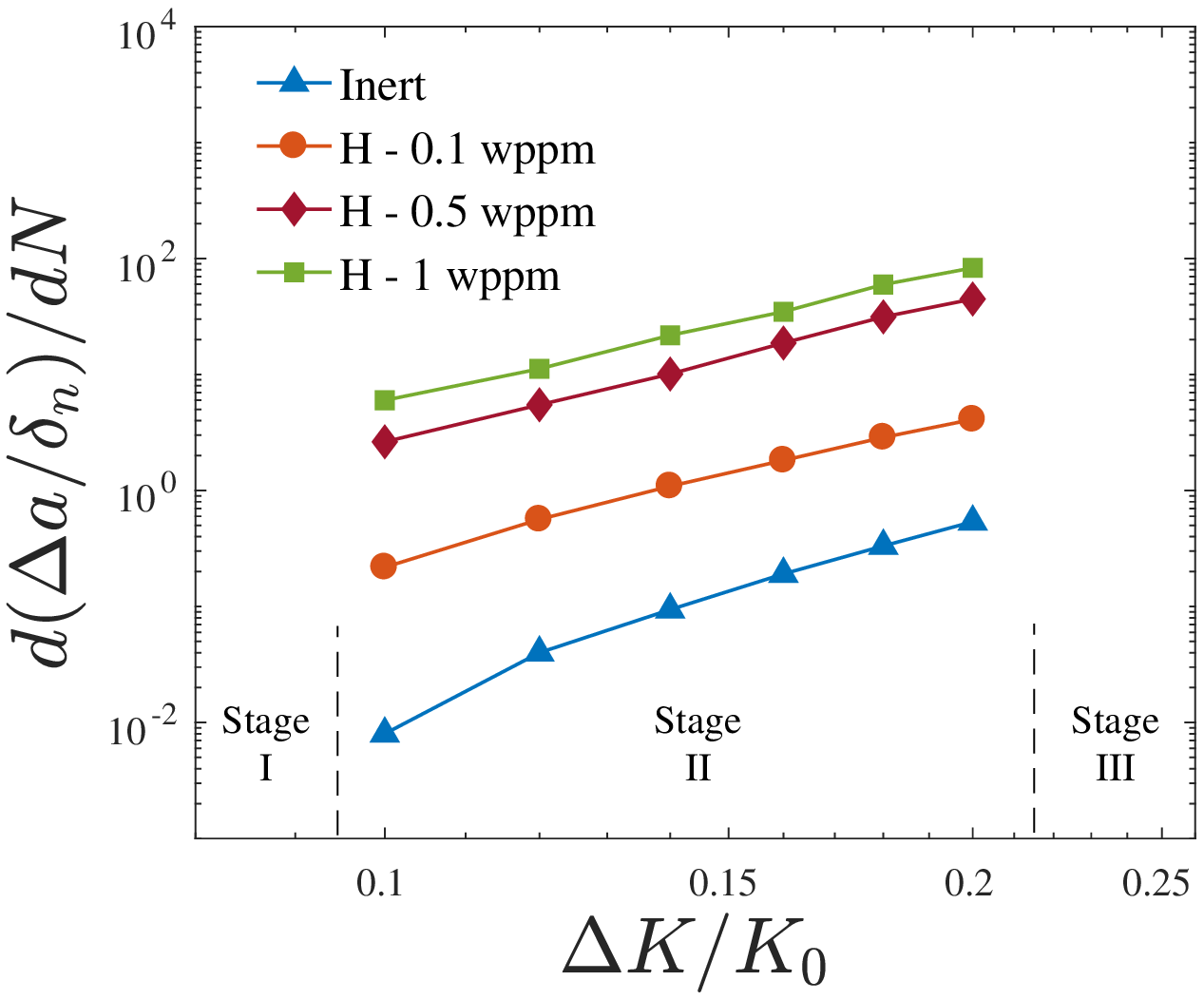}
                \caption{}
                \label{fig:InfluenceHb}
        \end{subfigure}
       
        \caption{Predicted influence of the environment on (a) crack extension versus number of cycles for $\Delta K/K_0=0.1$ and (b) fatigue crack growth rate versus load amplitude. Results have been obtained for an iron-based material under a load ratio of $R=0.1$ and frequency of $f=1$ Hz}\label{fig:InfluenceH}
\end{figure}

The capacity of the model to capture the sensitivity of fatigue crack growth rates to a hydrogenous environment is first investigated by computing the crack extension $\Delta a$ as a function of the number of cycles for different values of $C_L$ at the boundary. Figure \ref{fig:InfluenceH} shows the results obtained for a load ratio of $R=0.1$ and frequency of 1 Hz. The magnitude of the load ratio is appropriate for applications in the context of hydrogen-fueled vehicles, where load ratios between $R=0.1$ and $R=0.2$ accurately characterize the mechanical stresses resulting from filling cycles. Results reveal a strong influence of the environment, with crack propagation rates increasing significantly with the hydrogen content; the model appropriately captures the deleterious effect of hydrogen on crack growth resistance under cyclic loading conditions.\\

Lattice hydrogen concentrations at the boundary range from 1 wppm ($4.68 \cdot 10^{15}$ H atoms/mm$^3$), which corresponds to a 3\% NaCl aqueous solution \cite{G86}, to 0.1 wppm. The important role of hydrogen in the fatigue crack growth behavior can be clearly observed in the crack growth versus crack amplitude curves (Fig. \ref{fig:InfluenceHb}). By making use of the well-known Paris equation \cite{P61}:
\begin{equation}
\frac{da}{dN}=\mathcal{C} \Delta K^m
\end{equation}

\noindent one can easily see that $\mathcal{C}$ significantly increases with the environmental hydrogen content, in agreement with the experimental trends. On the other hand, results render a Paris exponent that shows little sensitivity to the hydrogen concentration, falling in all cases within the range of experimentally reported values for metals in inert environments ($m \approx 4$). The cycle-dependent contribution of the environment manifests significantly, while the influence of $C_b$ as $\Delta K$ increases is governed by a trade-off between larger levels of equivalent plastic strain (increasing $C_T$ and subsequently $C$) and shorter diffusion times due to greater crack growth rates. Thus, for a given frequency, the effect of hydrogen on the slope of the $da/dN$ versus $\Delta K$ curve depends heavily on the diffusion and mechanical properties of the material under consideration. The sensitivity of a steel to hydrogen embrittlement is therefore bounded between two limit cases: (a) \emph{slow tests}, where the testing time significantly exceeds the diffusion time for hydrogen within the specimen, and (b) \emph{fast tests}, where the testing time is much less than the diffusion time. In the former bound, hydrogen atoms accumulate in the fracture process zone and the distribution of lattice hydrogen concentration is governed by Eq. (\ref{eq:DISP}). For sufficiently rapid tests the initial (pre-charged) hydrogen concentration and the contributions from reversible microstructural traps dominate the response. As a consequence, experiments reveal a relevant increase in $da/dN$ with decreasing frequency until an upper bound is reached where the load-cycle duration is sufficient to allow hydrogen to diffuse and fully saturate the crack tip fracture process zone \cite{M12}.\\

\begin{figure}[H]
        \centering
        \begin{subfigure}[h]{1\textwidth}
                \centering
                \includegraphics[scale=0.8]{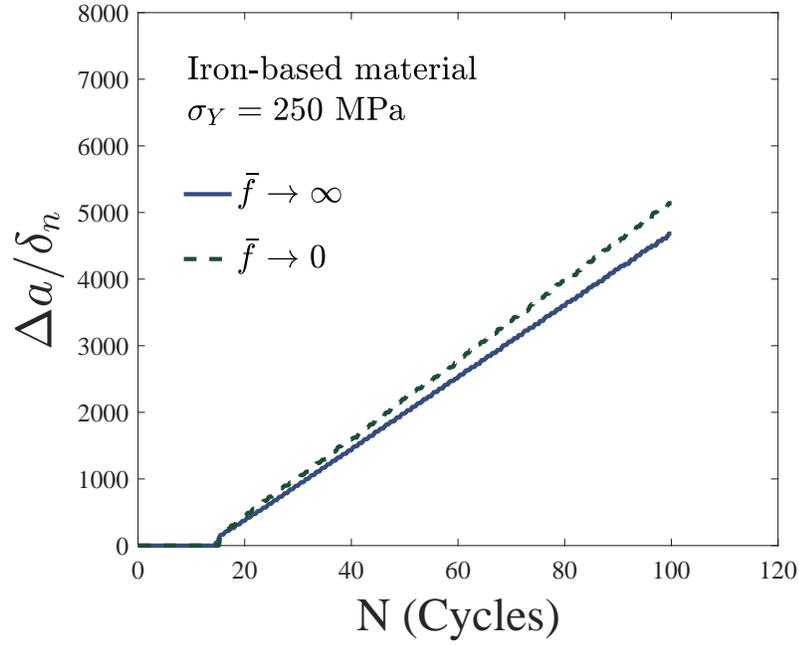}
                \caption{}
                \label{fig:InfluenceFreqIronA}
        \end{subfigure}
        
        \begin{subfigure}[h]{1\textwidth}
                \centering
                \includegraphics[scale=0.8]{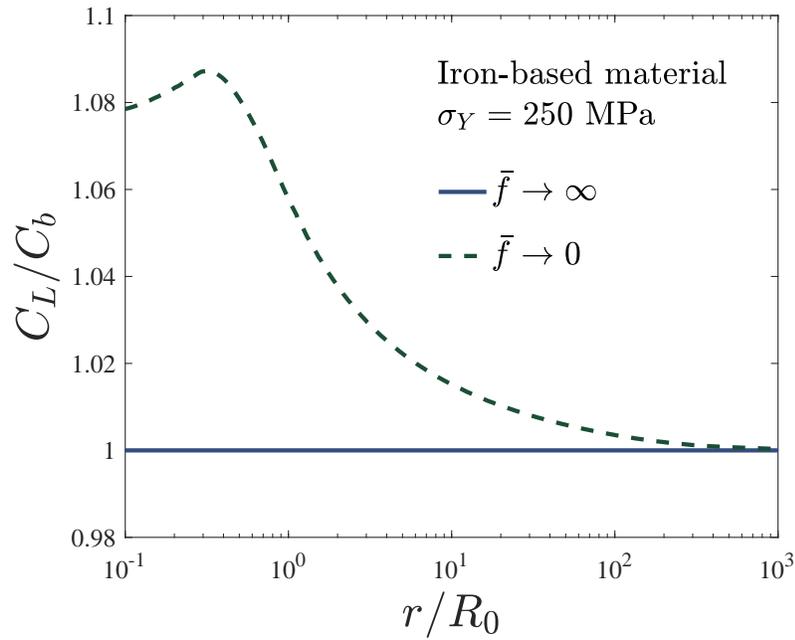}
                \caption{}
                \label{fig:InfluenceFreqIronB}
        \end{subfigure}
       
        \caption{Influence of the frequency in an iron-based material: (a) crack extension versus number of cycles, and (b) lattice hydrogen distribution ahead of the crack at the maximum $\Delta K$ and for a crack extension of $\Delta a /R_0=0.8$. Results have been obtained for $\Delta K/K_0=0.2$, under a load ratio of $R=0.1$ and an external hydrogen concentration of $C_b=1$ wppm.}\label{fig:InfluenceFreqIron}
\end{figure}

We subsequently investigate the influence of the loading frequency. A normalized frequency is defined as,
\begin{equation}
\bar{f}=\frac{f R_0^2}{\mathcal{D}}
\end{equation}

\noindent so as to quantify the competing influence of test and diffusion times. Fig. \ref{fig:InfluenceFreqIronA} shows crack growth resistance curves obtained for the iron-based material under consideration in the aforementioned asymptotic limits - \emph{slow tests} ($\bar{f} \to 0$) and \emph{fast tests} ($\bar{f} \to \infty$). In agreement with expectations, crack propagation is enhanced by larger testing times but results reveal very little sensitivity to the loading frequency, as opposed to experimental observations. Fig. \ref{fig:InfluenceFreqIronB} provides the basis for the understanding of the small susceptibility of crack growth rates to the loading frequency; the $C_L$ elevation in the $\sigma_H$-dominated case is less than 10\% of the lattice hydrogen concentration in the \emph{fast test}.\\

The difference between the two limiting cases is governed by the exponential dependence of the lattice hydrogen concentration to hydrostatic stresses ahead of the crack, given the independence of the trap density to the loading frequency in Sofronis and McMeeking's \cite{SM89} framework. Since the maximum level of $\sigma_H$ is load-independent in finite strain J2 plasticity \cite{M77}, we investigate the influence of yield strength, strain hardening and triaxiality conditions in providing a response closer to the experimental observations.\\

\begin{figure}[H]
        \centering
        \begin{subfigure}[h]{1\textwidth}
                \centering
                \includegraphics[scale=0.8]{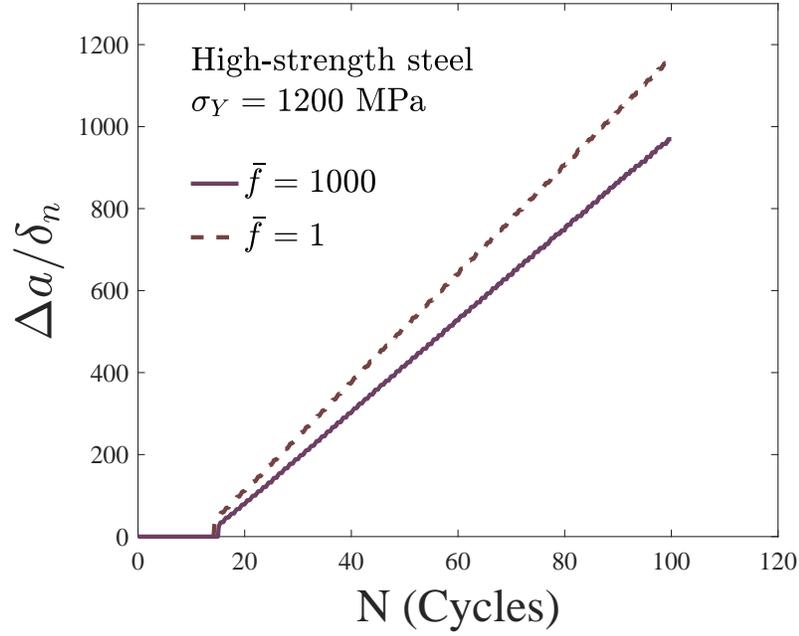}
                \caption{}
                \label{fig:InfluenceFreqHSA}
        \end{subfigure}
        
        \begin{subfigure}[h]{1\textwidth}
                \centering
                \includegraphics[scale=0.8]{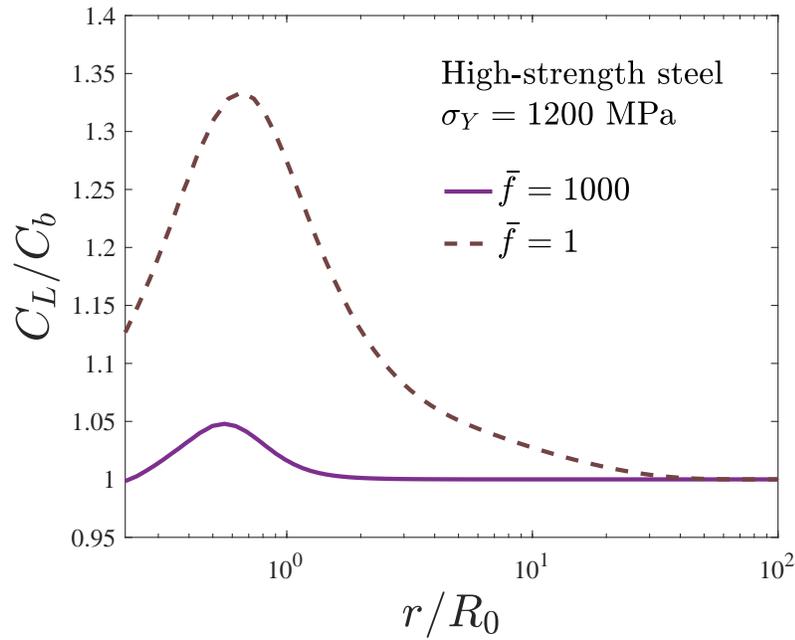}
                \caption{}
                \label{fig:InfluenceFreqHSB}
        \end{subfigure}
       
        \caption{Influence of the frequency in a high-strength steel: (a) crack extension versus number of cycles, and (b) lattice hydrogen distribution ahead of the crack at the maximum $\Delta K$ and for a crack extension of $\Delta a /R_0=0.6$. Results have been obtained for $\Delta K/K_0=0.2$, under a load ratio of $R=0.1$ and an external hydrogen concentration of $C_b=1$ wppm.}\label{fig:InfluenceFreqHstrength}
\end{figure}

The role of the yield strength is first investigated by considering a high-strength steel with $\sigma_Y=1200$ MPa and otherwise identical properties as the iron-based material assessed so far. As shown in Fig. \ref{fig:InfluenceFreqHSA}, a considerably larger effect of the loading frequency is observed, even without the need of considering the two limiting cases. The lattice hydrogen concentration ahead of the crack tip is shown in Fig. \ref{fig:InfluenceFreqHSB}; results reveal a much larger stress elevation when compared to the low-strength case (Fig. \ref{fig:InfluenceFreqIronB}).\\ 

\begin{figure}[H]
\centering
\includegraphics[scale=0.9]{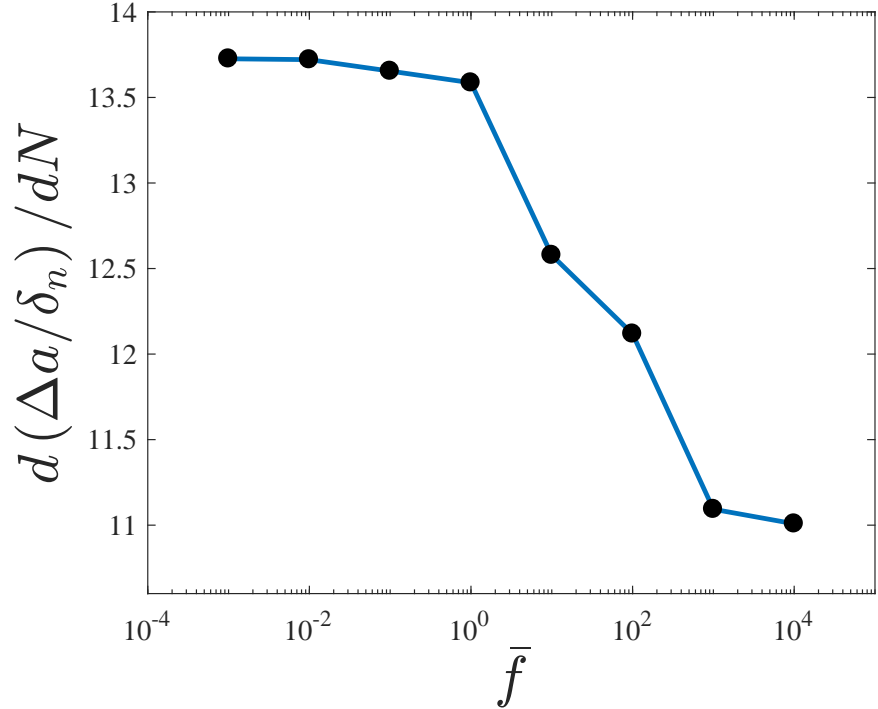}
\caption{Fatigue crack growth rate versus normalized frequency in a high-strength steel. Results have been obtained for $\Delta K/K_0=0.2$, under a load ratio of $R=0.1$ and an external hydrogen concentration of $C_b=1$ wppm.}
\label{fig:FreqHS}
\end{figure}

The increase in fatigue crack growth rates with decreasing frequency is quantified in Fig. \ref{fig:FreqHS}. Again, the model qualitatively captures the main experimental trends; low loading frequencies enable hydrogen transport to the fracture process zone, augmenting crack propagation rates. This $da/dN$-dependence with frequency reaches a plateau when approaching the two limiting responses, a clear transition between the upper and lower bounds can be observed. However, crack growth rates on the lower frequency bound are less than 1.5 times the values attained when $da/dN$ levels out at high loading frequencies; these quantitative estimations fall significantly short of reaching the experimentally reported differences. A 5-10 times crack growth rate elevation has been observed when decreasing frequency in a mid-strength martensitic SCM435 steel \cite{T07}, and similar data have been obtained for a 2.25Cr–1Mo (SA542-3) pressure vessel steel \cite{SR82} and an age-hardened 6061 aluminum alloy \cite{K07}, among many other (see, e.g., the review by Murakami \cite{M12}).\\

\begin{figure}[H]
\centering
\includegraphics[scale=0.9]{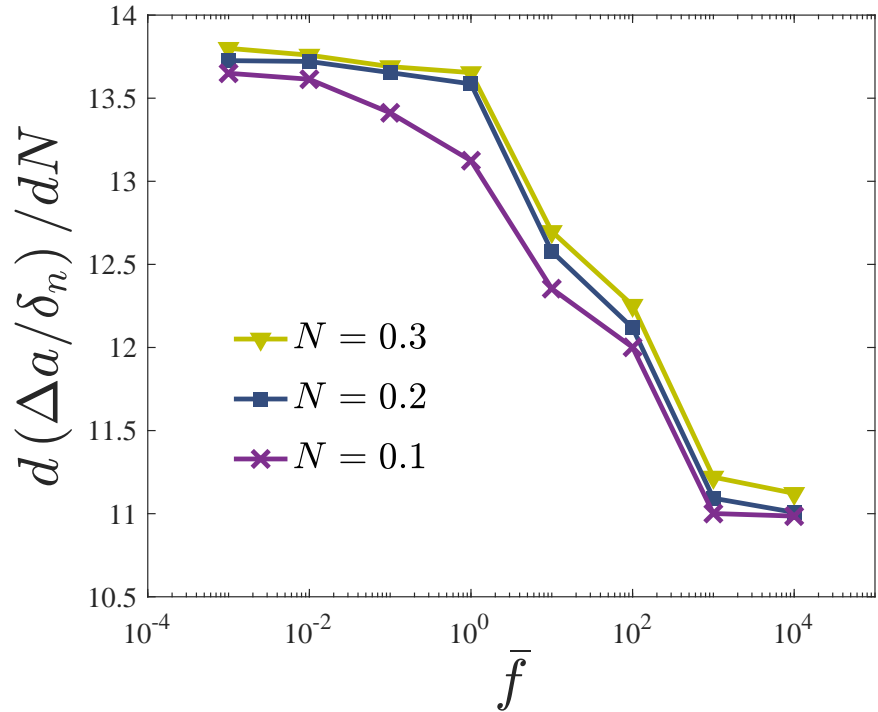}
\caption{Fatigue crack growth rate versus normalized frequency in high-strength steel for different strain hardening exponents. Results have been obtained for $\Delta K/K_0=0.2$, under a load ratio of $R=0.1$ and an external hydrogen concentration of $C_b=1$ wppm.}
\label{fig:N}
\end{figure}

The gap between the maximum and minimum $da/dN$ levels can also be affected by the strain hardening of the material under consideration. We, therefore, estimate the fatigue crack growth rates as a function of the loading frequency for three different strain hardening exponents. As shown in Fig. \ref{fig:N}, higher values of $\mathcal{N}$ lead to higher crack propagation rates. This comes as no surprise as larger strain hardening exponents translate into higher stresses. However, the $da/dN$-elevation is not very sensitive to the range of loading frequencies examined. The effect of the stress elevation due to larger $\mathcal{N}$ values could be attenuated by the intrinsic reduction of the plastic strain contribution to $N_T$. One should, however, note that, for the present choice of cohesive parameters, cracking takes place without significant plastic deformation. A different choice will probably increase the differences between the two limiting cases but highly unlikely to the level required to attain a quantitative agreement with the experiments.\\

Crack tip constraint conditions are also expected to play a role in augmenting crack growth rates in sufficiently slow tests. Here, we make use of the elastic $T$-stress \cite{BH91} to prescribe different triaxiality conditions by means of what is usually referred to as a modified boundary layer. Hence, the displacements at the remote boundary now read,
\begin{equation}
u \left( r, \theta \right) = K_I \frac{1+\nu}{E} \sqrt{\frac{r}{2 \pi}} \cos \left( \frac{\theta}{2} \right) \left(3 - 4 \nu - \cos \theta \right) + T \frac{1-\nu^2}{E} r \cos \theta
\end{equation}
\begin{equation}
v \left( r, \theta \right) = K_I \frac{1+\nu}{E} \sqrt{\frac{r}{2 \pi}} \sin \left( \frac{\theta}{2} \right) \left(3 - 4 \nu - \cos \theta \right) - T \frac{\nu (1+\nu)}{E}  r \sin \theta
\end{equation}

\begin{figure}[H]
\centering
\includegraphics[scale=0.9]{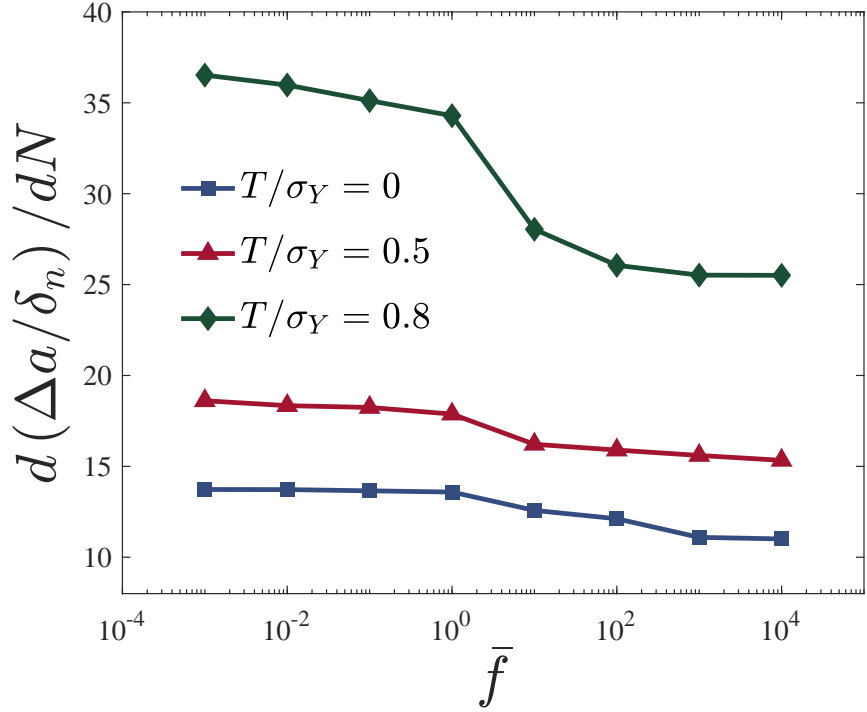}
\caption{Fatigue crack growth rate versus normalized frequency in high-strength steel for different constraint conditions. Results have been obtained for $\Delta K/K_0=0.2$, under a load ratio of $R=0.1$ and an external hydrogen concentration of $C_b=1$ wppm.}
\label{fig:Tstress}
\end{figure}

Fig. \ref{fig:Tstress} shows the sensitivity of $da/dN$ to the loading frequency under different constraint conditions. We restrict our attention to positive values of the $T$-stress, as lower triaxialities will not contribute to increasing crack growth rates in the lower frequency bound. Results reveal a substantial increase of $da/dN$ with increasing crack tip constraint. However, the influence on the ratio between the crack propagation rates for \emph{slow} and \emph{fast} tests is almost negligible.\\

Results provide a mechanistic interpretation of the reduction in fatigue crack resistance with decreasing frequency observed in the experiments. By properly incorporating the kinetics of hydrogen uptake into the fracture process zone, model predictions can be employed to identify the critical frequency above which the time per load cycle is insufficient for diffusible hydrogen to degrade the crack growth resistance. Accurate estimations are however hindered by the lack of quantitative agreement with experimental data regarding the impact of loading frequency on $da/dN$. Tests conducted at the low-frequency bound lead to crack growth rates that are 5-10 times larger than the values attained in experiments with a duration much shorter than the diffusion time. Such differences cannot be attained in the framework of conventional J2 plasticity, where the peak $\sigma_H$ (on the order of $5 \sigma_Y$) is insufficient to draw in sufficient levels of hydrogen to cause a 5-fold increase in $da/dN$. Crack growth rates at low loading frequencies increase with yield strength, material hardening and constraint conditions, but not even the most critical combination of these parameters appears to provide a quantitative agreement with a phenomenon that is observed in a wide range of metallic alloys and testing configurations. Additional sources of stress elevation are therefore needed to provide a reliable characterization of environmentally assisted fatigue for different frequencies. One possibility lies on the large gradients of plastic strain present in the vicinity of the crack, which exacerbate dislocation density and material strength. Geometrically necessary dislocations (GNDs) arise in large numbers to accommodate lattice curvature due to non-uniform plastic deformation, and act as obstacles to the motion of \emph{conventional} statistically stored dislocations. Strain gradient plasticity theories have been developed to extend plasticity theory to the small scales by incorporating this dislocation storage phenomenon that significantly contributes to material hardening (see \cite{M16c} and references therein). Gradient plasticity models have been consistently used to characterize behavior at the small scales involved in crack tip deformation, predicting a much higher stress level than classic plasticity formulations (see, e.g., \cite{MN16,M17}). This stress elevation due to dislocation hardening has proven to play a fundamental role in fatigue \cite{BS08,BS08b} and hydrogen-assisted cracking \cite{M16,MB17}.

\section{Conclusions}
\label{Concluding remarks}

We propose a predictive cohesive modeling framework for corrosion fatigue. The model is grounded on the mechanism of hydrogen embrittlement, which governs fatigue crack initiation and subsequent propagation in a wide range of metallic alloys exposed to gasses and electrolytes. Mechanical loading and hydrogen transport are coupled through lattice dilatation due to hydrostatic stresses and the generation of traps by plastic straining. An irreversible cohesive zone model is employed to capture material degradation and failure due to cyclic loads. The impact of the hydrogen coverage in the cohesive traction is established from first principles quantum mechanics. Finite element analysis of a propagating crack reveals a relevant increase in crack growth rates with (i) hydrogen content in the surrounding environment and (ii) decreasing load frequency; in agreement with experimental observations. A robust and appropriate numerical model for hydrogen-assisted fatigue opens up many possibilities, enabling rapid predictions that could be key to risk quantification in industrial components. Moreover, important insight can be gained into the mechanisms at play, identifying the relevant variables and their critical magnitudes for a given material, environment and loading scenario.\\

The influence of the yield strength, work hardening and constraint conditions is extensively investigated aiming to quantitatively reproduce the relation between the loading frequency and the crack growth rates observed in the experiments. Results reveal the need to incorporate additional sources of stress elevation to sufficiently enhance hydrogen uptake into the fracture process zone. Future work will focus on extending the present scheme to encompass the role of geometrically necessary dislocations through strain gradient plasticity formulations. 
 
\section{Acknowledgments}
\label{Acknowledge of funding}

The authors gratefully acknowledge financial support from the Ministry of Economy and Competitiveness of Spain through grant MAT2014-58738-C3. E. Mart\'{\i}nez-Pa\~neda additionally acknowledges financial support from the People Programme (Marie Curie Actions) of the European Union's Seventh Framework Programme (FP7/2007-2013) under REA grant agreement n$^{\circ}$ 609405 (COFUNDPostdocDTU).




\end{document}